\documentclass[aps,epsfig,twocolumn,showpacs]{revtex4-1}
\usepackage{amssymb}
\usepackage{graphics}
\usepackage{graphicx}
\usepackage{epsfig}
\begin{document}
\title{Effect of salt concentration on the stability of heterogeneous DNA}
\author{Amar Singh, Navin Singh}
\affiliation{Department of Physics, Birla Institute of Technology \& Science, Pilani - 333 031, 
Rajasthan, India}
\begin{abstract}
We study the role of cations on the stability of double stranded DNA (dsDNA) molecules.
It is known that the two strands of double stranded DNA(dsDNA) have negative
charge due to phosphate group. Cations in the form of salt in the solution, act as 
shielding agents thereby reducing the repulsion between these strands. We study
several heterogeneous DNA molecules. We calculate the phase diagrams for DNA molecules 
in thermal as well as in force ensembles using Peyrard-Bishop-Dauxois (PBD) model. 
The dissociation and the stacking energies are the two most important factors that 
play an important role in the DNA stability. With suitable modifications in the model 
parameters we investigate the role of cation concentration on the stability of different
heterogeneous DNA molecules. The objective of this work is to understand how these 
cations modify the strength of different pairs or bases along the strand. 
The phase diagram for the force ensemble case (a dsDNA is pulled from an end) 
is compared with the experimental results.
\end{abstract}
\maketitle

\section{Introduction}
The double helical structure of DNA molecule is stable due to the hydrogen bonding 
between the bases on the complimentary strands as well as due to the stacking interaction 
between the bases along the strand \cite{watson}. The double stranded state of 
these molecules strongly depends on the base pair composition, the sequence, and the ionic 
nature of the solution. The stability of double stranded DNA (dsDNA) can be studied either by 
changing the base pair composition, temperature, the pH of the solution (solvent),
or by applying the force on the chain \cite{rudolf,kumar,frank,hatch,ritort}. The 
experimental and theoretical findings have established that the melting temperature 
($T_m$) and the critical force (force required to unzip the chain, $F_c$) increases 
linearly with the $GC$ content of the DNA. It has also been established that the 
melting temperature increases non-linearly with the total salt concentration of the solution
\cite{owc,santa,Chaurasiya, weber}. This is because of the fact that the two strands 
of dsDNA are negatively charged (the negative charge on phosphate group). Hence, any change 
in the concentration of cations ($Na^+$ or $Mg^{2+}$) in the solution will affect the overall 
stability of DNA molecule \cite{amarphysica, krueger}. The addition of salt may neutralize 
the Coulombic repulsion between the phosphate groups of these two strands thereby stabilizing 
the molecule in double stranded state \cite{nik, manning, tomic}. However, the role 
of salt concentration on the mechanical unzipping or stretching is still not very clear. 
One should note that the response of the system (DNA) to the applied mechanical 
stress and to the thermal fluctuations is different. While in the case of thermal 
denaturation, opening is entropic in nature \cite{pb,sares,joyuex2008}, in case of mechanical 
stress it is enthalpic \cite{huguet,somen}. That is why the investigations on the role of salt 
concentration on the mechanical unzipping of DNA molecule are very important to understand 
the phenomenon of transition from close to open state in DNA. In most of the previous studies 
the salt effect or the ionic nature of the solvent were assumed to have a constant value 
\cite{ffalo,macedo}. Some semi-empirical calculations 
show that the thermal stability of the DNA molecule depends on the sequence heterogeneity 
of the molecule\cite{krueger}. However not much attention has been paid to understand 
the role of cations to the stability of heterogeneous DNA molecule that is stretched by a force.

In this manuscript, we investigate the role of cations on the critical force of heterogeneous
DNA molecule. To calibrate our model parameters we investigate the thermal denaturation of 
$\lambda$-phage DNA that is discussed in section \ref{temp}. The model Hamiltonian that is
used for the investigation has been discussed in section \ref{model}, where we also 
describe the method to calculate the melting temperature $T_m$, and the critical force $F_c$. 
The phase diagram for $\lambda$-phage DNA for mechanical unzipping case is discussed
in section \ref{force}. Section \ref{summ} summarizes the findings and discusses the
direction for future work.

\section{The model}
\label{model}
Theoretically, the DNA can be modelled like two polymeric chains that are inter-linked.
Various statistical mechanics based models have been proposed in the past which falls
broadly in two categories: Poland Scheraga (PS) model \cite{poland,yeramian,hanke} 
which consider dsDNA chain with regions of denaturated loops, and Peyrard Bishop Dauxois 
(PBD) model \cite{pb,pey2009} which is a Hamiltonian based model. Some theoretical models 
assume DNA as two self avoiding or directed walks on a square lattice 
\cite{kumar,somen,kumar_epl} and study the thermal as well as mechanical denaturation
of dsDNA molecule. For the current investigation we use PBD model which considers the stretching 
between corresponding bases only. Although the model ignores the helicoidal structure of the 
dsDNA molecule, it has enough details to analyze mechanical behavior at few ${\rm \AA}$ scale 
relevant to molecular-biological events\cite{ns3,ns4}. The Hamiltonian for DNA, containing
$N$ base pairs, is written as,
\begin{equation}
\label{eqn1}
H = \sum_{i=1}^N\left[\frac{p_i^2}{2m} + W_S(y_i,y_{i+1}) + V_M(y_i) \right]
\end{equation}
where $y_i$ represents the stretching from the equilibrium position of the hydrogen bonds. 
The term $p_i = m\dot{y}_i$ represents the momentum while $m$ is the reduced mass of a base pair, 
taken to be the same for both $AT$ and $GC$ base pairs. 
The stacking interaction between two consecutive base pairs along the chain 
is represented by, 
\begin{equation}
\label{eqn2}
W_S(y_i,y_{i+1}) = \frac{k}{2}(y_i - y_{i+1})^2[1 + \rho e^{-b(y_i + y_{i+1})}],
\end{equation}
where $k$ represents the single strand elasticity. The anharmonicity in the strand 
elasticity is represented by $\rho$ while $b$ represents its range. The stacking interaction
$W_S(y_i,y_{i+1})$ is independent of the nature of the bases at site $i$ and $i+1$ as 
these parameters are assumed to be independent of sequence heterogeneity. However, the sequence 
heterogeneity has effect on the stacking interaction along the strand. This can be taken 
care through the single strand elasticity parameter $k$. 

The hydrogen bonding between the two bases in the $i^{\rm th}$ pair is represented by 
the Morse potential $V_M (y_i) = D_i(e^{-a_iy_i} - 1)^2$. The on-site potential is modified 
to incorporate the solution effect as \cite{amarphysica, macedo},
\begin{equation}
\label{eqn3}
V_M(y_i) = D_i(e^{-a_iy_i} - 1)^2 -\frac{1}{4}D_i\left[\tanh\left(\gamma y_i\right)-1\right] 
\end{equation}
where $D_i$ represents the potential depth, roughly equal to the bond energy of that pair 
and $a_i$ represents the inverse of the width of the potential well. The
heterogeneity in the base pair sequence is taken care by the values of $D_i$ and $a_i$. 
An additional term in the Morse potential is the solvent term which simulates the formation 
of hydrogen bonds with the solvent, once the hydrogen bonds are stretched by more than 
their equilibrium values. The \textquotedblleft$\tanh$\textquotedblright term in the potential 
enhances the energy of the equilibrium configuration and the height of the barrier below 
which the base pair is closed \cite{amarphysica,zoli,ffalo_pre}. 

In the stability of the dsDNA molecule the role of hydrogen bond is the key factor. 
In most of the previous studies, the hydrogen bond interaction and the effects of surroundings, 
such as ionic solution environment, are taken as constant. In our previous study \cite{amarphysica} 
we have modified the potential depth to incorporate the salt effect as,
\begin{equation}
\label{eqn4}
D_i = D_0\left[1 + \lambda_1 \ln \left(\frac{C}{C_0}\right)+ 
\lambda_2 \ln^2 \left(\frac{C}{C_0}\right)\right]
\end{equation}
Here, the concentration, $C$ is expressed in moles per liter and $C_0$ is the reference 
concentration chosen to be 1 mole/liter. The $\lambda_i^s$ appearing in the 
potential are solution constants. The melting temperature, $T_m$, of the chain increases 
linearly with the $GC$ content of the DNA at a fixed salt concentration and non-linearly 
with the logarithm of the salt concentration at a fixed base pair composition.
As the bonding nature of $A-T$ is different to the $G-C$, the response of any change
in the salt concentration may be different for these base pairs, it would be interesting
to study the variable $\lambda_i^s$ for $AT$ and $GC$ base pairs and its effect on the
transition from double stranded to single stranded state.

Thermodynamics of the transition can be investigated by evaluating the expression for 
the partition function. For a sequence of $N$ base pairs with periodic boundary conditions, 
partition function can be written in terms of Hamiltonian as \cite{ns3}:
\begin{equation}
\label{eqn5}
Z = \int \prod_{i=1}^{N}\left\{dy_idp_i\exp[-\beta H(y_i,y_{i+1})]\right\} = 
Z_pZ_c,
\end{equation}
where $Z_p$ corresponds to the momentum part of the partition function while the $Z_c$
contributes as the configurational part of the partition function. Since 
the momentum part is decoupled in the integration, it can be integrated out as a 
simple Gaussian integral. This will contribute a factor of $(2\pi mk_BT)^{N/2}$ in the 
partition function, where $N$ is the number of base pairs in the chain. The configurational 
partition function, $Z_c$, is defined as \cite{ns3},
\begin{equation}
\label{eqn6}
Z_c = \int \prod_{i=1}^N dy_i  K(y_i,y_{i+1})
\end{equation}
where $K(y_i,y_{i+1}) = \exp\left[-\beta H(y_i,y_{i+1})\right].$
For the homogeneous chain, one can evaluate the partition function by transfer integral (TI)
method by applying the periodic boundary condition. In case of heterogeneous chain, with open 
boundary, the configurational part of the partition function can be integrated numerically with 
the help of matrix multiplication method. Once the  limit of integration has been chosen, the 
task is reduced to discretized the space to evaluate the integral numerically. We choose 
the limits as $-5.0 \; {\rm \AA}\; {\rm to}\; 200.0 \; {\rm \AA}$, as the lower and upper 
limits of the integration, respectively \cite{zhang}. The space is being discretized using 
the Gaussian quadrature formula with number of grid points equal to 900 \cite{ns3}. 
As all matrices in eq.\ref{eqn6} are identical in nature the multiplication is done 
very efficiently. The thermodynamic quantities of interest can be calculated by evaluating 
the Helmholtz free energy of the system. The free energy per base pair is,
\begin{equation}
\label{eqn7}
f(T) = -\frac{1}{2}k_B T\ln\left(2\pi m k_B T\right) - \frac{k_B T}{N}\ln Z_c.
\end{equation}
The thermodynamic quantities like specific heat ($C_v$), in the thermal or force
ensemble, can be evaluated by taking the second derivative of the free energy.
The peak in the specific heat corresponds to the melting temperature or the critical
force of the system \cite{amarphysica,ns3}.

\section{Thermal denaturation}
\label{temp}

Here we discuss the role of sequence heterogeneity on the temperature induced transition in the DNA molecule. 
The hydrogen bonding between bases of complementary strands and the stacking between 
neighboring bases stabilize the double-stranded structure of DNA  molecule. These base stacking 
interactions are of the order of magnitude of a few $k_BT$. Thermal fluctuations can 
cause the disruption of base pair bonds which ultimately leads to the unzipping of double helix. 
This is known as the {\it thermal melting} of dsDNA \cite{skumarysingh, alex}. Not only the bond energies 
of base pairs but the stiffness of the chain also contributes to the melting or denaturation 
of the dsDNA molecule \cite{zoli2014}. In the PBD model, the hydrogen bonding between the 
bases in a pair is represented by the Morse potential while the strand elasticity (or stiffness) 
is represented by the stiffness parameter $k$.
\begin{figure}[htb]
\label{fig01}
\begin{center}
\includegraphics[height=3.in,width=3.5in]{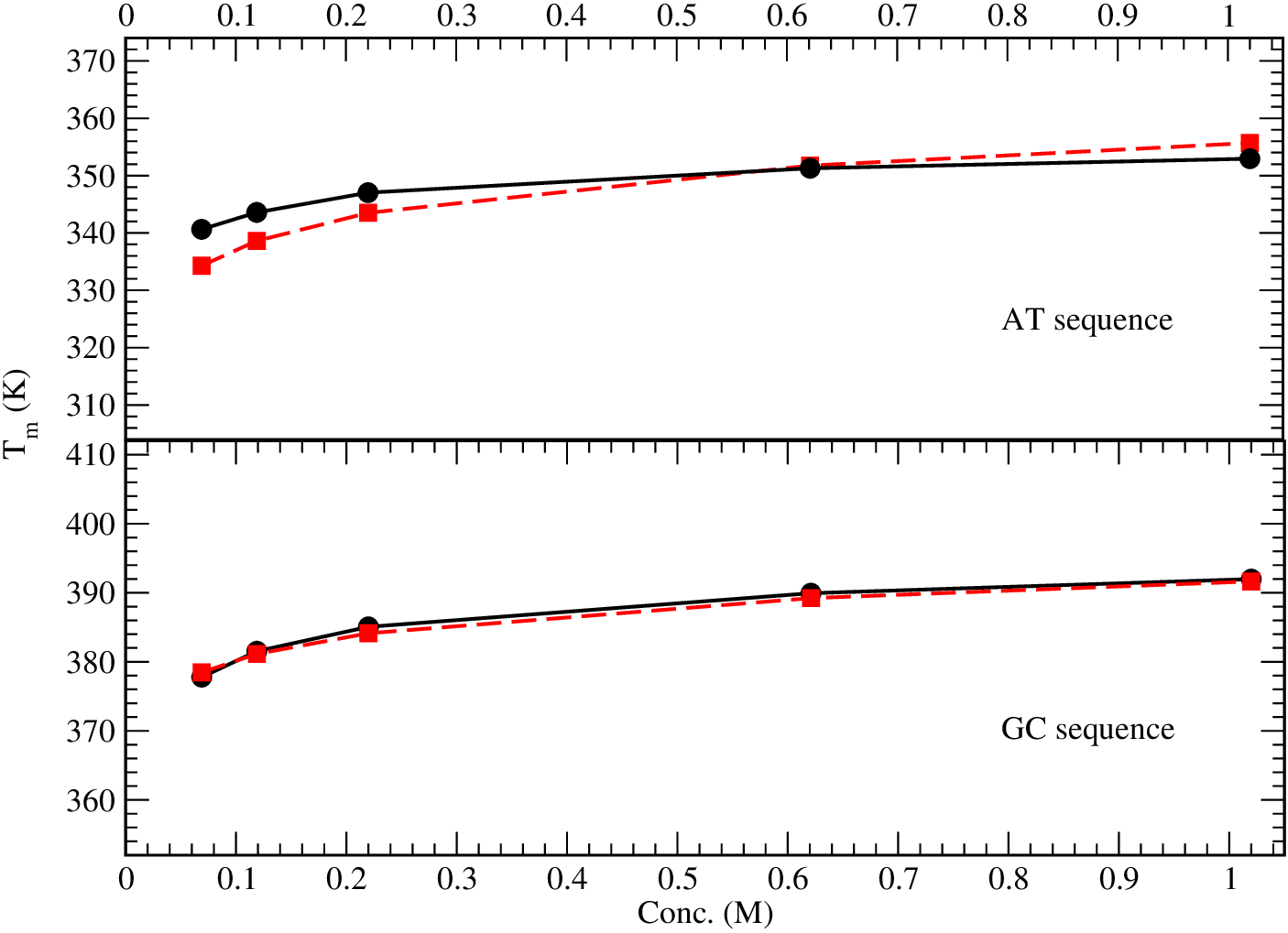}
\caption{The phase diagram calculated using the PBD model (black circle) and
from the empirical relation given by Krueger {\it et al} \cite{krueger} (red square).}
\end{center}
\end{figure}

In most of the thermal denaturation studies the chain stiffness parameter is taken as
constant. In this work, we consider the stiffness parameter, $k$, as well as the solution 
constants, $\lambda_1$ \& $\lambda_2$ as sequence dependent, {\it i.e.}, their values 
depend on the nature of base or base pair in the sequence. This means that the shielding due 
to cations will depend whether, along the sequence, there is an $AT$ or a $GC$ base pair. 
We calculate the melting temperature of the homogeneous $AT$ chain and homogeneous $GC$ 
chain to calibrate our results. First, we consider two kinds of homogeneous sequence
of 1000 base pairs, one that is having all $AT$ pairs and the other that is having all 
$GC$ base pairs. In fig. 1 we show the nature of denaturation of $AT$ and $GC$ chains and 
their dependence on the concentration of salt in the solution. We compare our results with 
the empirical calculations by Krueger {\it et al} \cite{krueger}. For pure $GC$ chain our 
model based calculations are in good match with Krueger results. However, there is a 
difference in the behaviour of pure $AT$ chain at lower salt concentration.
\begin{figure}[htb]
\begin{center}
\includegraphics[height=2.25in,width=3.5in]{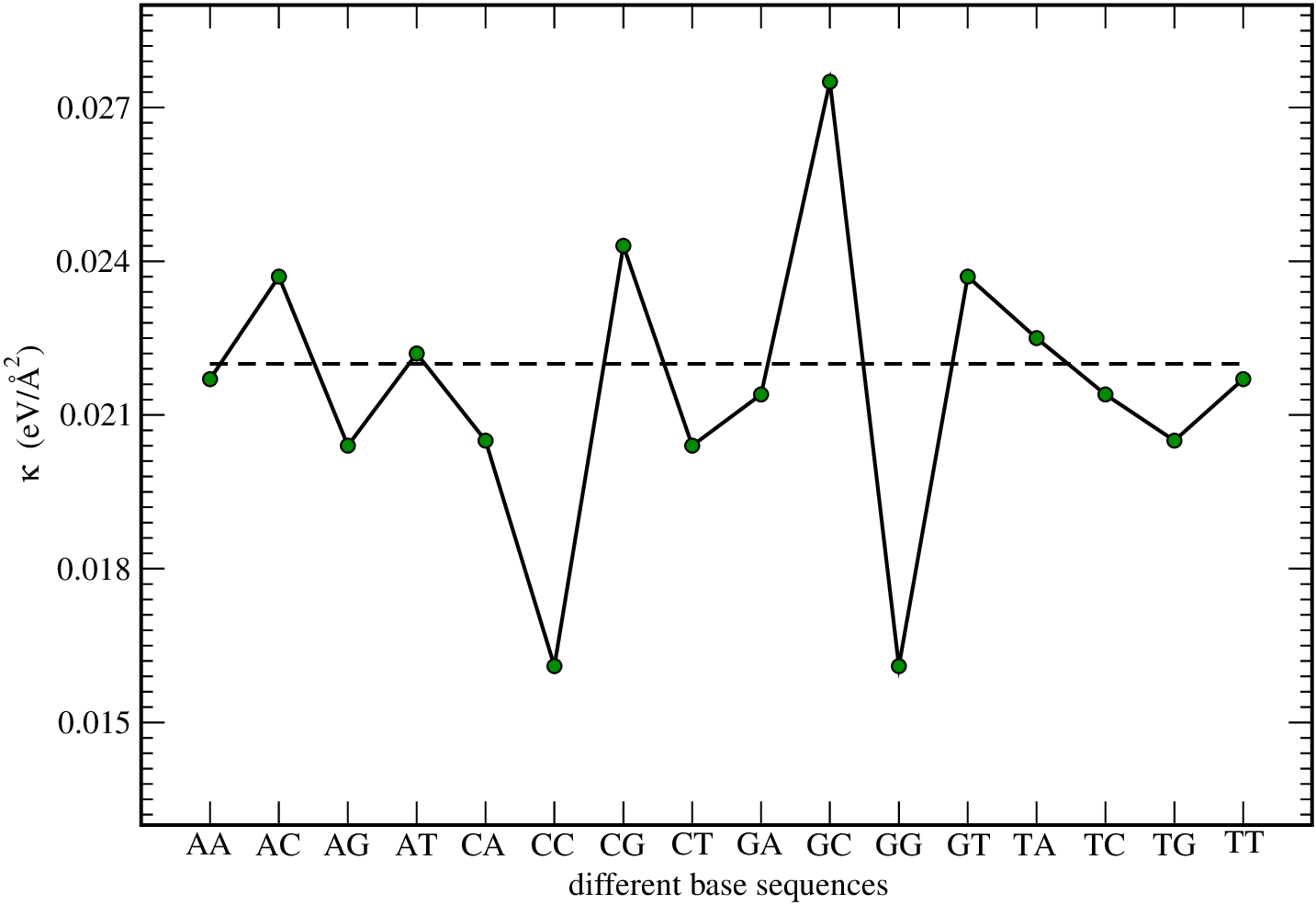}
\caption{Values of all 16 different stacking constants \cite{sponer} are shown. 
The stacking constant appears to be strongest for the $GC$ pair while it is showing 
the lowest value for $GG/CC$ stacking.}
\end{center}
\label{fig01}
\end{figure}

To investigate the sequence effect of salt concentration on the stability of DNA molecule,
we consider a chain of 1000 base pairs. The respective location of $AT$ \& $GC$ pairs along the
chain is random. We consider a chain that is having 100 $AT$ and 100 $GC$ pairs in a block
and these blocks appear in the sequence alternatively (chain 1). We consider another chain of 1000
base pairs that is a segment of $\lambda$ phase DNA (chain 2) \cite{sanger}. In the PBD model, the base
sequence is taken care by the model parameters, like potential depth, $D$, and the elastic 
constant, $k$. These parameters depend on the nature of pairing as well as on the sequence
of these bases along the chain. As the approximate ratio of $GC$ and $AT$ bond strengths
is 1.5, we take $D_0 (GC) = 1.5 D_0(AT)$. In the stacking energy term, the elastic constant 
$k$ is taken from refs. \cite{alex,sponer}. We take the average of the values mentioned 
in the refs. \cite{alex,sponer} and for each base we calculate the change in elastic energy 
and scaled it to the elastic constant $k_j$, where, $j$ is the index for any of the 
16 possible combination as shown in fig. 2. We adopt the following scheme to label the elastic constant
as a function of base sequence. Wherever, $AG$ stacking is mentioned, it means the base sequence
will be from $5'-A-G-3'$  and the reverse sequence will be on the complementary strand from $3' - 5'$. 
That is why we have only 10 unique stacking energies \cite{alex,sponer,cooper} (fig. 2).
The value of solution constant $\lambda$ is also different for $AT$ and $GC$ pairs. 
For our investigation, we consider four different cases:
(a) constant $k$ with $\lambda_{AT} = \lambda_{GC} $, (b) constant $k$ with $\lambda_{AT} 
\neq \lambda_{GC} $, (c) variable $k$ with $\lambda_{AT} = \lambda_{GC} $, 
(d) variable $k$ with $\lambda_{AT} \neq \lambda_{GC} $. The lowest value of the stacking
constant is for $GG$ and $CC$ stacking while it is highest for $GC$ stacking.
For cases (a) \& (b) the elastic constant is the average value of the variable $k$ 
and it is 0.022 ${\rm eV/\AA^2}$.

\begin{figure}[htb]
\begin{center}
\vspace{0.5cm}
\includegraphics[height=2.5in,width=3.0in]{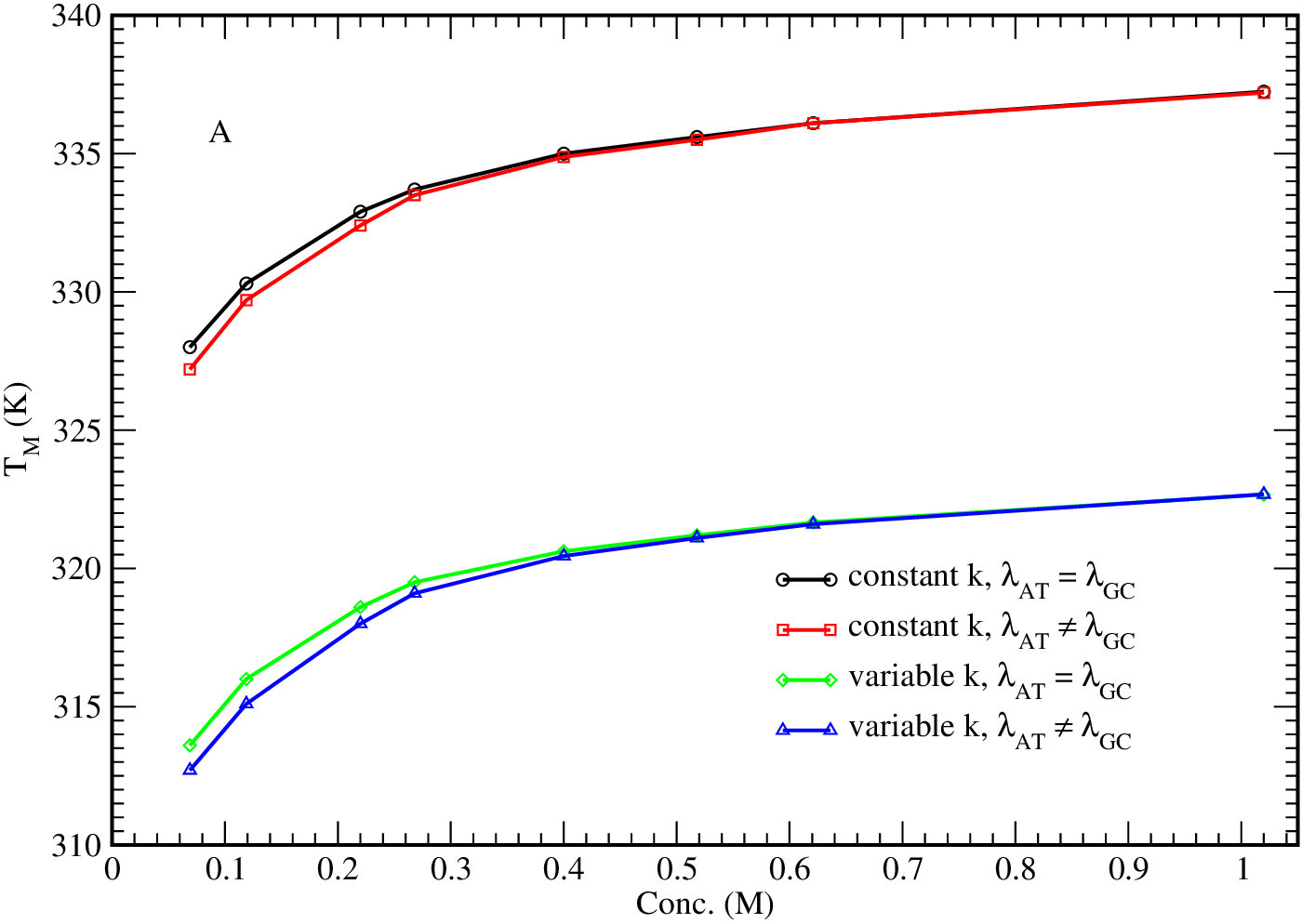}\hspace{0.25cm}
\includegraphics[height=2.5in,width=3.0in]{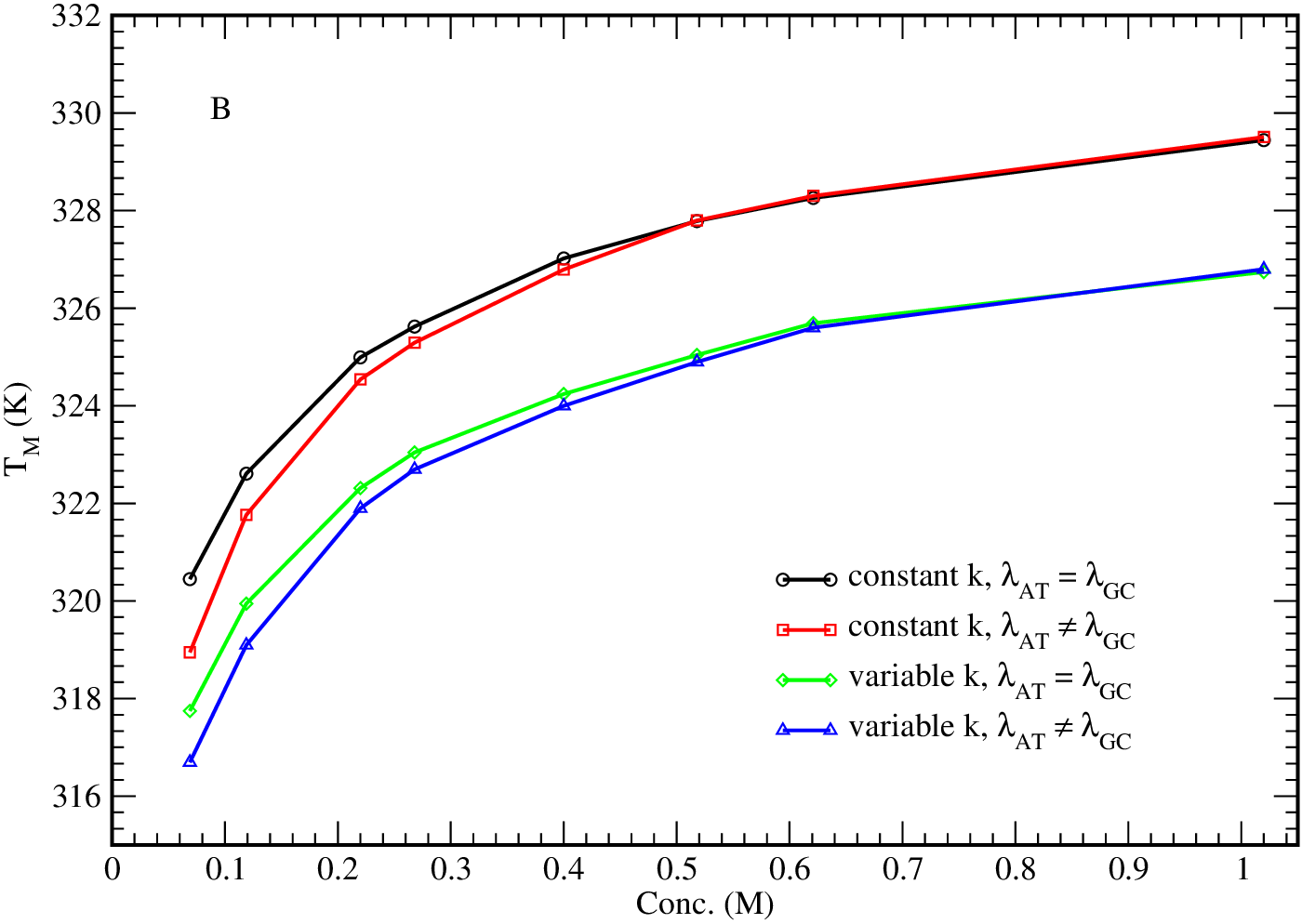}
\caption{The phase diagram for temperature induced transition in DNA. 
Figure 1A is for a chain that is having alternate $AT$ \& $GC$ pairs while figure
1B is for a section of $\lambda$-phage DNA.}
\label{fig03} 
\end{center}
\end{figure}
The results obtained for all the four cases are shown in the figure 3. From the 
figure this is clear that when we take constant stiffness parameter ($k$), the melting temperature increases 
by $\sim10$ K for any amount of salt in the solution. However, the solution constants 
$\lambda$ have significant effect as far as sequence dependence is concerned. If we compare 
either for constant $k$ (case(b)) or for variable $k$ (case(d)), the solution constant $\lambda$ 
affect the melting transition in the lower concentration range. While for higher concentrations, 
there is no significant change in melting temperature, in the lower range of concentration, 
not only the $T_m$ values ($\Delta T_m \approx 1$ K) are different but also the slope of the 
curve is different. The nature of the curves at lower concentrations indicates the difference 
in the opening or denaturation of different heterogeneous chains. This is an important point 
that was missing in the previous studies 
\cite{amarphysica,krueger}.

\section{Forced induced transition}
\label{force}
In this section, we investigate the force induced transitions [for all the (a)-(d) cases] 
in DNA that is surrounded by the cations. All the calculations are done at room temperature,
{\it i.e.} 300 K. The modified Hamiltonian for the force that is applied at either of the 
ends, is,
\begin{equation}
\label{eqn9}
H = \sum_{i=1}^N\left[\frac{p_i^2}{2m} + W_S(y_i,y_{i+1}) + V_M(y_i) \right] - F.y_e
\end{equation}
The force that is required to unzip the chain, the critical force ($F_c$),
is calculated in the constant force ensemble (CFE). Addition of the force term in the Hamiltonian
make the matrices, that appear in the equation of the partition function, assymetric. Hence the
transfer integral method is no longer valid. Here we multiply the matrices to get the free energy
of the system as a function of force \cite{zhang,cocco}.
We consider two kinds of heterogeneous chains as discussed above. 
For the forced induced unzipping investigations we take the elastic constant, $k$, as
0.022  ${\rm eV/\AA^2}$. The other important parameters like the potential depth, $D$,
the solution constants, $\lambda_i$ are tuned in order to get a good match with the experimental 
results. The set of values for which our results are found in better agreement with 
experiments are: $D = 0.064$ eV and $\lambda_{AT}$ = 0.012 \& -0.0024; 
$\lambda_{GC}$ = 0.008 \& -0.0016 \cite{lambda}. The complete set of parameters are listed in
Table 1. The melting temperature of the 1000 bps chain is around 319 K and 330 K for 
0.030 M \& 1.020 M respectively with these set of parameters. 
\begin{table}[ht]
\label{tab2}
\begin{center}
\begin{tabular}{|l|l|} \hline
{\bf Parameters} & {\bf Values} \\ \hline
Potential Depth, $D$ & 0.064 eV (AT) \& 0.096 eV (GC)\\ \hline 
Inverse potential width, $a$ & 4.2 (AT) \& 6.3 (GC) \\ \hline
Anharmonicity, $\rho$ & 1.0 \\ \hline  
Anharmonic Range, $b$ & 0.35 ${\rm \AA^{-1}}$ \\ \hline
Solution constant, $\lambda_1$ & 0.012 (AT) \& 0.008 (GC) \\ \hline
Solution constant, $\lambda_2$ & -0.0024 (AT) \& -0.0016 (GC) \\ \hline
Solution constant, $\gamma$ & 1.0 \\ \hline
\end{tabular}
\end{center}
\end{table} 
\begin{figure}[hbt]
\begin{center}
\vspace{0.5cm}
\includegraphics[height=2.25in,width=3.0in]{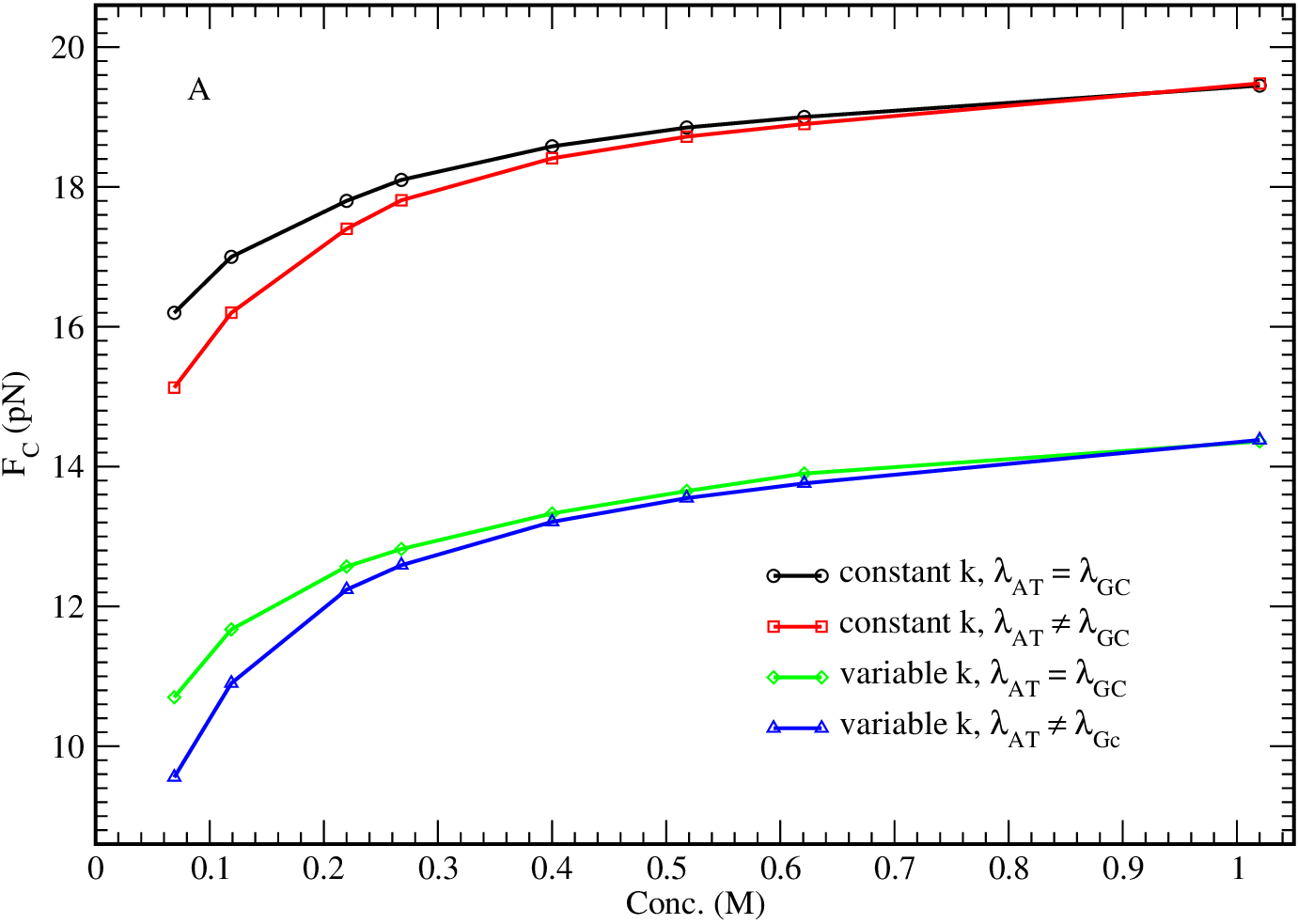}\hspace{0.35cm}
\includegraphics[height=2.25in,width=3.0in]{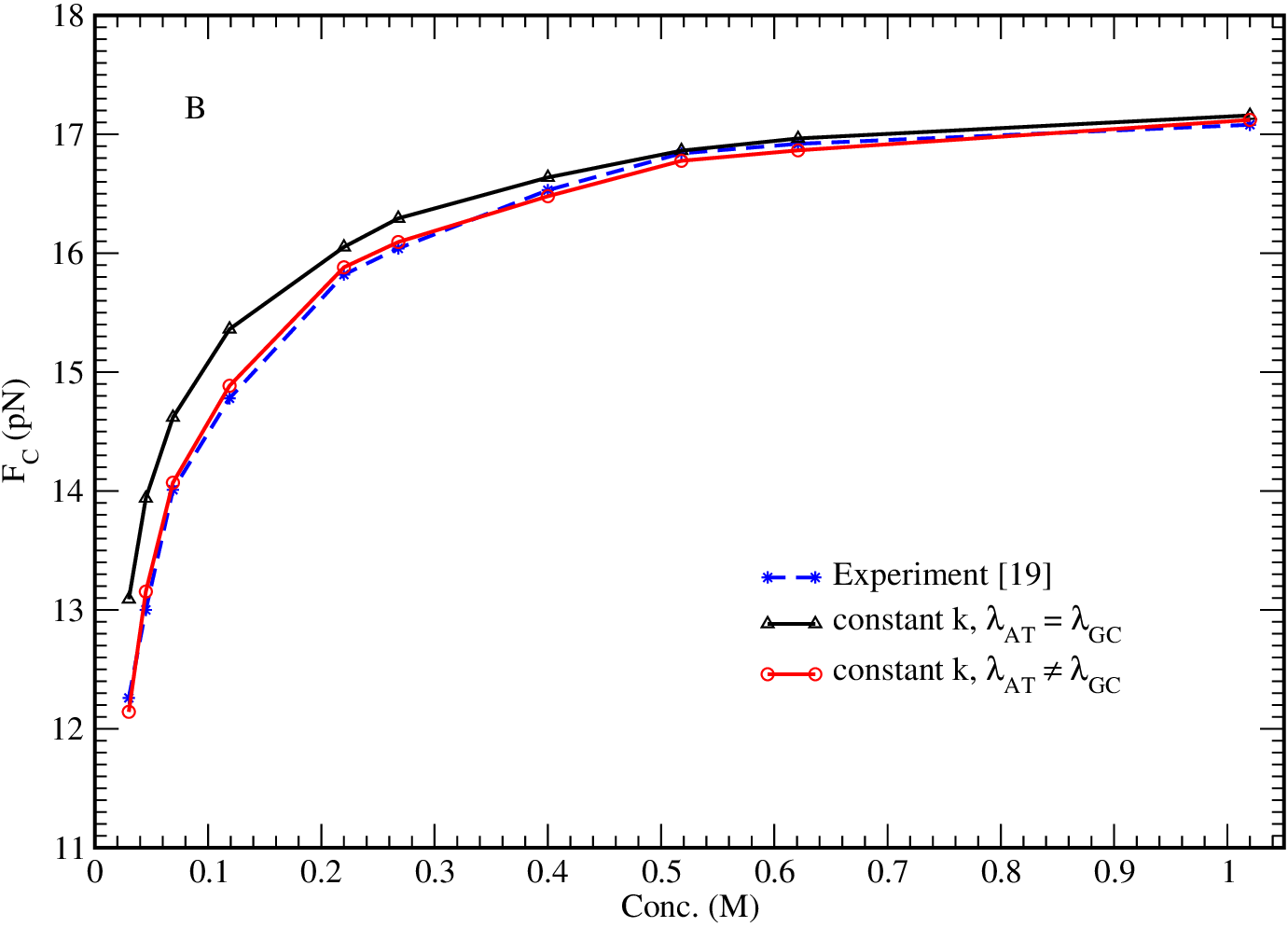}
\caption{The phase diagram of DNA chain when the DNA is pulled from one of the ends. 
Figure 2A is for a chain that is having alternate $AT$ \& $GC$ pairs while figure
2B is for a section of $\lambda$-phage DNA. Here we compare our results with the 
experimental result that is found by Huguet {\it et al} \cite{huguet}.}
\label{fig04} 
\end{center}
\end{figure}
The transition from close to open state is different for mechanical unzipping and
temperature denaturation. While in the temperature induced transition, the opening is 
due to formation of loops or bubbles in addition to the end entropy, in force (mechanical) 
induced transition the opening is primarily due to stretching of hydrogen bond that causes 
a pair to break  \cite{pkmaiti, rajeevkapri}. In this case, there is an interface 
of open and close state which forces system to move from one state to another depending 
on the value of applied force \cite{ns3}. We consider the four different cases 
as mentioned in sec. \ref{temp} in order to get more details of the force induced transition 
in DNA. In the force induced transition, we observe that there is smaller contribution
of variable stiffness parameter, $k$. We find that the solution constant $\lambda_i$ has more 
prominent effect on the phase diagram (as shown in fig. \ref{fig04}). For variable $\lambda_i$ (the solution
constant) at lower concentration the phase diagram shows the difference in its role on
$AT$ and $GC$ base pair (fig. 4). However at concentrations higher than 0.06 M, the sequence effect diminishes. 
When we take constant $k$ and variable $\lambda_i$ [case (c)], the obtained phase
diagram, in fig. 4B, shows a good match with the experimental phase diagram obtained 
by Huguet {\it et al} \cite{huguet}. The earlier result \cite{amarphysica} that showed some mismatch at
the lower concentrations is showing better match by including the sequence dependent salt
concentration. 
\begin{figure}[hbt]
\label{fig05} 
\begin{center}
\includegraphics[height=2.5in,width=3.5in]{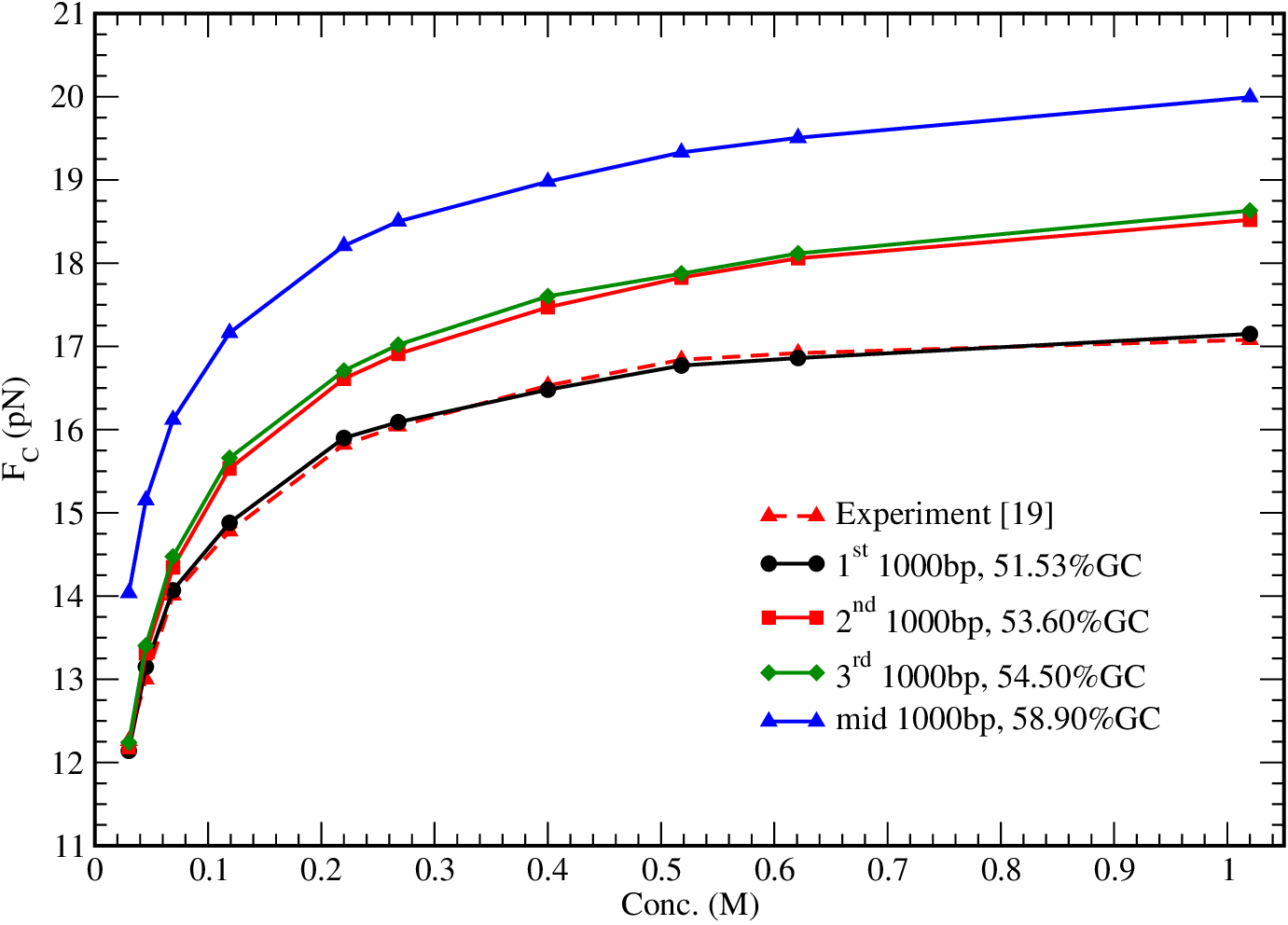}
\caption{The phase diagram of $\lambda$-phage DNA molecule that shows the variation in the 
critical force with salt concentration of the solution. Here we take different segments of the chain 
that varies not only in \%GC content but also in the sequence of the bases along the strand.
Again we compare our results with the experimental findings of Huguet {\it et al} \cite{huguet}.}
\end{center}
\end{figure}
To get more insight of the sequence dependent salt effect 
on the force induced transition in DNA, we take different segments of $\lambda$-phage DNA 
\cite{sanger} that is having different $GC$ content. The sequences that we take, are 
(i) $1^{st}$ 1000 bp having 51.30\%, 
(ii) $2^{nd}$ 1000 bp having 53.60\%, 
(iii) $3^{rd}$ 1000 bp having 54.50\% and 
(iv) a segment of 1000 bp anywhere from the sequence that is having 58.90\% of $GC$ pairs. 
We observe some interesting features of the transition from close to open state for different
sequences. The critical force for different DNA sequence varies as the $GC$ content varies.
However, the best match we are getting for the chain that is 
having $51.53\%$ of GC content. This is in accordance with the fact that real sequences of 
DNA is having about 50\% of $GC/AT$ pairs randomly distributed along the chain. 

\section{Conclusions}
\label{summ}
We have investigated the role of sequence dependent salt concentration on the stability of 
DNA molecule. As the interaction between $A-T$ and $G-C$ pairs are not same, the interaction
of cations with these these molecules may not be the same. We have modified the on-site potential
to take care the heterogeneity of the sequence ({\it i.e.} the presence of $AT$ or $GC$ pair). 
Using PBD model, we have investigated the role of cations on the melting temperature as well
as on the critical force that is required to unzip the chain from one end. 
In the mechanical unzipping case, we have considered different chains that are having 
different $GC$ content. Our results are in good match with the experimental results 
for the chain that is having 51\% $GC$ content. This is in accordance with the fact that
for the experiments, in general, the sequence that is used, are having $\sim 50\%\; GC$ content.
The sequence heterogeneity can be introduced by variable bond strength (between the pairs) 
and by choosing variable stacking energies for possible 16 conformations of the base stacking
in the PBD model. Hence, we chose variable values of chain stiffness $k$ as well as the solution 
constants $\lambda_i$ to analyze the sequence effect on the stability of the molecule.  
We found that the role of cations, that are interacting with different pairs, on the stability 
of the molecule ($D_i$ \& $\lambda_i$) have more impact than the variable base stacking
$k_j$. This may be due to fact that the hydrogen bonding play more crucial role than the 
stacking energy in the overall stability of the molecule. However, this should be noted
here that our calculations are based on the average value of stacking as well as bond 
energies. The role of $Na^+$ or $Mg^{2+}$ in the overall stability would be interesting 
to study.

\section*{Acknowledgement}
We are thankful to Yashwant Singh and Sanjay Kumar, Department of Physics, Banaras Hindu
University, India, for useful discussions. We acknowledge the financial support provided by 
University Grant Commission, New Delhi, India and Department of Science and Technology,
New Delhi under the different project grants and BSR fellowships.


\begin{thebibliography}{100}
\bibitem{lambda}
We have chosen these particular values of $\lambda$, because the average of $\lambda_{AT}$ \&  
$\lambda_{GC} $ is equal to the values of $\lambda$ in Ref. \cite{amarphysica}.
\end{thebibliography}
\end{document}